\documentclass{article}

\usepackage{arxiv}

\usepackage[authoryear,longnamesfirst]{natbib}
\usepackage[utf8]{inputenc} 
\usepackage[T1]{fontenc}    
\usepackage{hyperref}       
\usepackage{url}            
\usepackage{booktabs}       
\usepackage{amsfonts}       
\usepackage{nicefrac}       
\usepackage{microtype}      
\usepackage{lipsum}
\usepackage{graphicx}
\usepackage{booktabs}
\usepackage{amsmath}
\usepackage{algorithm,algpseudocode}
\usepackage{makecell}
\usepackage{mathtools}
\usepackage{float}

\newcommand\blfootnote[1]{%
  \begingroup
  \renewcommand\thefootnote{}\footnote{#1}%
  \addtocounter{footnote}{-1}%
  \endgroup
}

\graphicspath{ {./images/} }

\DeclareUnicodeCharacter{2212}{-}

\title{Influence Maximization in Social Networks using Discretized Harris' Hawks Optimization Algorithm and Neighbor Scout Strategy}

\author{
 Inder Khatri\footnotemark[1] \\
  Biometric Research Laboratory\\
  Delhi Technological University\\
  New Delhi, India\\
  \texttt{inderkhatri999@gmail.com} \\
  \And
 Arjun Choudhry\footnotemark[1] \\
  Biometric Research Laboratory\\
  Delhi Technological University\\
  New Delhi, India\\
  \texttt{choudhry.arjun@gmail.com} \\
  \And
 Aryaman Rao\footnotemark[1] \\
  Department of Electrical Engineering\\
  Delhi Technological University\\
  New Delhi, India\\
  \texttt{aryaman26601@gmail.com} \\
  \And
 Aryan Tyagi \\
  Department of Mechanical Engineering\\
  Delhi Technological University\\
  New Delhi, India\\
  \texttt{tyagiaryan82@gmail.com} \\  
  \And
  Dinesh Kumar Vishwakarma \\
  Biometric Research Laboratory\\
  Delhi Technological University\\
  New Delhi, India\\
  \texttt{dinesh@dtu.ac.in} \\
  \And
  Mukesh Prasad \\
  School of Computer Science \\
  University of Technology Sydney\\
  Ultimo, Australia\\
  \texttt{mukesh.prasad@uts.edu.au} \\  
}

\begin{document}
\maketitle

\begin{abstract}
Influence Maximization (IM) is the task of determining $k$ optimal influential nodes in a social network to maximize the influence spread using a propagation model. IM is a prominent problem for viral marketing, and helps significantly in social media advertising. However, developing effective algorithms with minimal time complexity for real-world social networks still remains a challenge. While traditional heuristic approaches have been applied for IM, they often result in minimal performance gains over the computationally expensive Greedy-based and Reverse Influence Sampling-based approaches. In this paper, we propose the discretization of the nature-inspired Harris' Hawks Optimisation meta-heuristic algorithm using community structures for optimal selection of seed nodes for influence spread. In addition to Harris' Hawks' intelligence, we employ a neighbour scout strategy algorithm to avoid blindness and enhance the searching ability of the hawks. Further, we use a candidate nodes-based random population initialization approach, and these candidate nodes aid in accelerating the convergence process for the entire populace. We evaluate the efficacy of our proposed DHHO approach on six social networks using the Independent Cascade model for information diffusion. We observe that DHHO is comparable or better than competing meta-heuristic approaches for Influence Maximization across five metrics, and performs noticeably better than competing heuristic approaches.\blfootnote{*Equal Contribution}
\end{abstract}


\section{Introduction}

Over the years, Social Networks (SNs) have gained immense popularity in our modern-day society. Large Social Networks like Twitter, Facebook, and Instagram, which have billions of users and connections, serve as viable communication tools as well as important marketing platforms for advertisers due to their significant user base. Recent studies \citep{recentStudies1,recentStudies2,recentStudies3} show that people tend to believe and trust the recommendations by individuals in their social circles, rather than the ads from more conventional means like television advertisements, billboards, etc. This ‘word-of-mouth’ approach tends to make information have a cascading effect to reach a wider range of people, gradually increasing the oral publicity effect. Utilizing this concept, advertisers choose a set of initial influencers and make them promote their product through ‘word-of-mouth’ to produce the largest cascade spread possible and influence a wide demographic of people on social networks. This approach is commonly known as viral marketing. Therefore, choosing the optimal initial users so that the information is propagated as widely in the network as possible is the core objective of influence maximization as a task.

Influence Maximization (IM) can be formally defined as the task of identifying the optimal set of nodes in a network that, when activated, can maximize the spread of information over the entire social network. \citet{Domingos2001} were the first group of researchers who studied IM. They introduced a novel solution by modeling the network as a Markov random field. \citet{Kempe2003} officially defined the problem of IM, proving that, when simulated using Independent Cascade (IC) and Linear Threshold (LT) models, IM behaves like an NP-hard problem. They further introduced a hill-climbing strategy-based greedy approach to identify optimal initial spreader nodes for maximum information spread. Their approach was evaluated for hundreds of Monte-Carlo simulations for every node, making it computationally expensive. This seriously restricts its usability in larger networks. 

To minimize the computational cost and space complexity, the optimality of IM had to be compromised. Later, certain approaches were proposed to meet the required purpose of less complexity and great accuracy. \citet{Leskovec2007} presented \textit{Cost-Efficient Lazy Forward (CELF)}, an enhanced greedy approach that reduced the computational cost of the greedy algorithm by approximately 700 times. \citet{Chen2009} further introduced two novel algorithms called NewGreedy and MixGreedy, which brought substantial improvements over \citet{Kempe2003}'s greedy approach. While NewGreedy and MixGreedy brought substantial computational cost reductions as compared to the approach introduced by \citet{Kempe2003}, they still were unviable for large networks.

\citet{jiang2011simulated} proposed the Expected Diffusion Value (EDV) approach, which was used to approximate the spread of influence in the Independent Cascade model, calculated using Monte Carlo simulations. They further used a simulated annealing-based approach to optimize EDV, which was found to give substantially better results with millions of iterations. Their experiments showed that the EDV algorithm is more accurate and 100 to 1000 times faster than the greedy algorithm. \citet{simulatedAnnealing} further proposed an alternative method for Simulated Annealing and two heuristic algorithms. They showed that their simulated annealing algorithm led to significantly faster convergence, making their approach more applicable to larger networks.

Some recent approaches for IM have also incorporated Community Structures \citep{CBIM, Neuro1, CAOM, CFIN}, and depend on the proposition that a social network can be divided into groups based upon how many communities can be made in that network. A Community structure is a group of interconnected in close vicinity. A closed and clustered community is generally easier to completely influence than one which is open and sparsely connected. It is much easier to spread influence rapidly in a community with only a few seed individuals. These community structures are found to be more efficient than other centrality measure-based approaches and can be applied more efficiently to larger social networks.
In this work, we propose the discretization of Harris' Hawks Optimization (DHHO) algorithm with community structures for Influence Maximization. We employ the Louvain algorithm for detecting significant communities in the networks, which are further used by the DHHO algorithm to detect optimal nodes for maximum influence spread. We also introduce a novel neighbor scout strategy and an adapted randomized population initialization approach for improved performance. We evaluate DHHO against seven novel approaches on five metrics across six datasets. Experimental results verify DHHO performs comparably or better than competing meta-heuristic approaches for IM. DHHO further outperforms competing heuristic and centrality measures-based approaches. Our contributions are summarized below.
\begin{itemize}
    \item We introduce DHHO, a discretized Harris' hawks optimization algorithm to optimize the Local Influence Estimator (LIE) function for Influence Maximization, using the intelligence of the hawk population. We propose the approach to maximize the expected influence spread by a set of extracted seed nodes, across any given online social network. We have employed a random population initialization strategy, along with a novel neighbor scout strategy to accelerate the process of convergence and achieve further improved results.

    \item We employ community structures for narrowing down the search space for our DHHO algorithm to find optimal candidate nodes, thus increasing the efficiency of our approach as well as reducing the execution time. Community clusters aid in information propagation within the community and reduce the time complexity significantly. We detect community structures in social networks using the Louvain Algorithm and assign a seed node budget to each significant community on the basis of its size and degree centrality. Further, a candidate pool of optimal nodes is created for the purpose of influence maximization using the DHHO algorithm.
    
    \item We evaluate DHHO on six real-world social networks and compare it with seven baseline approaches across five different performance metrics. Our experimental results verify that DHHO outperforms competing meta-heuristic approaches, as well as the other novel heuristics algorithms and centrality measures. Statistical tests further verify our results, proving the superiority of DHHO over competing approaches.
\end{itemize}

The remaining paper is structured as follows: Section 2 delineates the existing research works distributed by the type of algorithm employed by the said approaches. Section 3 contains the preliminaries needed for our approach. Section 4 contains our proposed approach and algorithms. Section 5 contains information about various experimental details like the datasets used, the performance metrics, and the implementation environment. Section 6 consists of our experimental results, their analysis, and broad outcomes. Section 7 elucidates our statistical tests to evaluate the performance of our approach. Section 8 contains our concluding statement.

\section{Related Works}

Since \citet{Domingos2001} treated Influence Maximization as an algorithmic task using Markov Random Field Theory, various approaches have been introduced for Influence Maximization in social networks. These can be broadly classified into four classes, which have been discussed below in significant detail.

\subsection{Greedy Algorithms-based Approaches}

Following \citet{Domingos2001}, \citet{Kempe2003} showed that Influence Maximization can be simulated using information diffusion models like Linear Threshold (LT), Independent Cascade (IC), and Weighted Independent Cascade (WIC), implying that the task of selecting the most influential nodes in a network is an NP-hard problem. Thus, \citet{Kempe2003} proposed a hill-climbing search strategy-based greedy algorithm on the sub-modular function to find an optimal solution within a factor of $1-1/(e-\epsilon)$. Here, e represents the natural logarithm, while $\epsilon$ is the error obtained due to the Monte Carlo simulation. One major drawback of greedy algorithms is that they need to choose the best $k$ seed nodes in the network iteratively. They select a single seed node at a time, and then the remaining nodes in the network need to be traversed. Thus, tens of thousands of Monte-Carlo simulations are required to accurately assess the spread of influence due to a given seed node set. This makes the greedy algorithm computationally expensive. 

The high time complexity of greedy algorithms severely limits their ability to deal with large social networks, and thus they are primarily used for small networks with lesser nodes. To overcome these barriers, \citet{Leskovec2007}, by exploiting the sub-modularity property, stored nodes with maximum marginal gain in a priority queue, and introduced the Cost-Effective Lazy-Forward (CELF) algorithm for finding the optimal seed nodes more efficiently. CELF was found to achieve solution accuracy similar to traditional greedy algorithms, while reducing the number of Monte-Carlo simulations by nearly 700 times when evaluating the marginal gain of nodes. \citet{Goyal2011} later proposed the CELF++ strategy to reduce the time complexity of CELF, and was found to be up to 55\% faster than CELF. 

\citet{Chen2009} introduced two novel algorithms called NewGreedy and MixGreedy. The NewGreedy algorithm reduces the time complexity of the algorithm by scraping those edges from the graph which don’t contribute to the dissemination of information from the original network to construct a smaller network. The MixGreedy algorithm, on the other hand, integrates the ideas from NewGreedy and CELF to further boost the efficiency of the algorithm. \citet{Cheng2013} later presented StaticGreedy, a new greedy algorithm that stored the static information of Monte-Carlo simulations within the network, decreasing the Monte-Carlo simulations by about two orders of the total magnitude. 

\citet{Heidari2015} proposed a machine greedy algorithm, that evaluates the magnitude of traversing nodes during the estimation propagation process. They further constructed a Monte Carlo graph in the simulation process. \citet{Lu2016} presented a novel approach to evaluate the spread of influence in the network recursively through the utilization of the reachability probability between the nodes. By applying the greedy strategy, nodes with the highest estimate were recursively selected as seed nodes in the network. Contrary to more conventional greedy algorithms, the computational cost of these algorithms is greatly diminished. However, these approaches are still computationally expensive, especially for larger networks.

\subsection{Reverse Influence Sampling-Based Algorithms}

\citet{Borgs2012} introduced Reverse Influence Sampling (RIS) for the selection of influential nodes in a network, an approach derived from the random sampling theory. It generates $1 − 1/(e − \epsilon)$ viable solutions with a computational cost of $(m + n)\epsilon−3logn$. RIS works on the fundamental idea that if a node frequently occurs in the randomly generated sub-graphs, then it has a higher probability of being a seed node. \citet{Tang2014} later proposed TIM and TIM+ algorithms to improve upon the RIS algorithm with greater algorithmic efficiency and significantly lesser time complexity by introducing improved sampling mechanisms. \citet{Tang2015} later introduced IMM, a novel approach derived from the Martingale estimation. IMM generates similar optimal solutions as TIM+ and TIM while having significantly higher efficiency, thus mitigating their drawbacks.

\citet{Nguyen2016} recently introduced SSA and D-SSA, two RIS-based algorithmic approaches, with 1200 times the computational efficiency of IMM. They further adhere to the same $1 − 1/(e − \epsilon)$ approximation guarantee. While the IMM algorithm has nearly linear time complexity, its practical sampling time noticeably exceeds the threshold for the minimum sampling time in actual operation. Since SSA and D-SSA are derived from the stop-and-stare strategy, they alleviate the shortcomings of IMM. Both of these approaches exhibited the same approximate optimal result guarantee as that of IMM but with around 1200 times lesser execution time.

\citet{Cohen2014} developed SKIM, a new greedy Sketch-based algorithm based on building combined reachability sketches. While it maintains the same approximation guarantee as that of TIM+, it can easily be applied to extremely large networks with billions of nodes and edges. Some other sketch-based RIS frameworks that were proposed are BKRIS \citep{Wang2017} and RSRS \citep{Kim2017}. They were introduced to ameliorate the drawbacks of SKIM and TIM+. While RIS-based approaches typically exhibit lesser computational costs as compared to greedy approaches, they sometimes return sub-optimal results on increasing the number of seeds.

\subsection{Heuristics Methods}

Recent works on IM have introduced a variety of heuristic approaches involving node centrality \citep{heuristics1,heuristics2}. These involve a simple protocol to explicitly choose the top $k$ nodes out of the set of total nodes when ranked according to the given centrality methods. Some prominent Heuristic approaches are Closeness Centrality \citep{CC}, Betweenness centrality \citep{Estrada2005}, Degree centrality \citep{Freeman1978}, and PageRank \citep{Brin1998}.

Over the years, more advanced heuristic algorithms have been introduced by researchers. \citet{kshell} proposed K-shell decomposition, a global centrality measure, which divides the nodes into different levels on the basis of their location with respect to the core of the network. \citet{Kundu2011} proposed the Diffusion Degree algorithm using the two-hop neighbourhood region to gauge a node's influence. \citet{ENC} proposed an improved algorithm called Neighbourhood Coreness measure, which ranks nodes on the basis of the sum of the K-shell of the neighbourhood nodes. This concept was further extended to get Extended Neighbourhood Coreness by taking the sum of the Neighbourhood Coreness values of a node's neighbours. \citet{MDD} introduced Mixed Degree Decomposition (MDD), an iteration of K-shell that further distinguishes nodes in the same shell by using the removed and residual neighbours during the K-shell decomposition process. Inspired by the concept of gravity, \citet{Ma2016} introduced GC-Index (gravity centrality index) for IM. Centrality-measures-based heuristic algorithms aid in the selection of influential nodes efficiently. However, they sometimes suffer from low accuracy and stability.

Community structures often hide subtly in online social networks. Nodes belonging to the same community are often closely connected with each other. Consequently, nodes from different communities have relatively sparse connections \citep{Newman2006}. Thus, information propagation is much easier within a community structure \citep{Lin2015}. Some of the most relevant community-based IM approaches introduced in the last few years are CAOM \citep{CAOM}, CFIN \citep{CFIN}, and GLR \citep{GLR}. The community-based works focus on finding out the core nodes in the significant communities, and further leverage the higher connection density of communities to increase the spread using these nodes as initial spreaders. 

\citet{Chen2014} tackled IM using community structures in the heat propagation model. They introduced the CIM approach, which performed very well for IM. However, its efficacy in other common information diffusion models remains to be tested. While IM approaches based on community structures are reasonably efficient, they frequently use infection propagation of nodes amongst their own communities to approximate the node influence in the whole social network. Thus, their accuracy is critically dependent on the validity of the community partitions.

\subsection{Metaheuristic Methods}
Recently, researchers have applied evolutionary optimization algorithms for IM. For these, a fitness function needs to be defined. \citet{jiang2011simulated} introduced the Expected Diffusion Value (EDV), which is shown in Eq. \ref{Eq:EDV}.
\begin{equation}
    EDV(S) = k + \sum_{v\epsilon N_S^{(1)} \backslash S}(1-(1-p)^{r(v)})
    \label{Eq:EDV}
\end{equation}
Here, $N_S^{(1)}$ signifies the one-hop area of the candidate set $S$, $N_S^{(1)}$ signifies the nodes in the one-hop area except for the candidate nodes $S$, $r(v)$ is the number of edges connected to node $v$ in $N_S^{(1)}$, and $p$ represents the activation probability.

\citet{jiang2011simulated} further adopted the simulated annealing approach to identify a seed set $S$ that maximizes the EDV fitness function. Their experiments showed that the algorithm is more accurate and 100 to 1000 times faster as compared to the greedy algorithm. 

\citet{Gong2016} introduced a Local Influence Estimator (LIE) function, which evaluates a set $S$'s influence on its neighbours within a two-hop area. They introduced a Discrete Particle Swarm Optimization (DPSO) framework to maximize the LIE fitness function. We delineate the LIE 
function in Section 3.4. \citet{Cui2018} proposed the Degree-Descending Search Evolution (DDSE) algorithm for IM. It utilises the EDV function to overcome the inefficiencies shown by the greedy algorithm. They underscored the reduced computational cost of DDSE as compared to greedy approaches in their experimental results. \citet{Zareie2020} used the Gray Wolf optimization algorithm for IM, along with a new fitness function inspired by the concept of entropy.

The Discrete Bat Optimization algorithm for IM was introduced by \citet{batAlgo}. \citet{Ma2016} pointed out that the LIE fitness function ignores interactions between neighbours in the same-hop surroundings from the seed node. They introduced a three-layer influence evaluation approach and adopted an improved bee colony optimization framework for the maximization of the three-layer evaluation model. The biggest advantage of using meta-heuristic algorithms is that running the Monte-Carlo simulations can be avoided. This increases efficiency and reduces the time complexity of the problem. However, it is crucial that the evolution mechanism chosen is effective in producing a high-quality and high-efficiency solution.

\section{Preliminary Information}

\subsection{Influence Maximization}
IM, as defined by \citet{Kempe2003}, is the process of choosing $k$ initial seed nodes in a social network graph G, such that when these nodes are considered as influential spreader nodes, the information diffusion over the entire network is maximum for any association of $N$ users \citep{Kempe2003}. The seed set $S$ of most influential nodes contains $k$ elements $(k \leq n)$. The influence spread from a set of seed nodes $S$ is maximum at the end of the diffusion process and is denoted as $\sigma (S)$. The aim is to increase information diffusion and expand the set of nodes influenced. Generally, the value of $k$ is minuscule as compared to $N$, considering the assets and time constraints. Mathematically, IM can be defined as in Eq. \ref{Eq:IM}.
\begin{equation}
    S= \arg\max \sigma(S)\ where\
    S \subset V, |S| = k
    \label{Eq:IM}
\end{equation}

\subsection{Information Diffusion Models}

Information Diffusion (ID) can be defined as the process of propagating information in a social network \citep{IP2, IP1}. It aids in evaluating various spreader-ranking methods. To model information propagation, usually, epidemic models are used. In a network, nodes act as the population, while the information propagation is represented as an infectious disease traversing through the population on interaction. The Linear Threshold (LT) \citep{LT}, Weighted Independent Cascade (WIC) \citep{WIC}, and Independent Cascade (IC) models \citep{IC} are some of the most common diffusion models used to simulate information propagation through a network. We employ the Independent Cascade model for simulating information diffusion in our evaluations. 

In the Independent Cascade model, every node is either in an infected (or active) state or in a susceptible (or inactive) state. An active node would be a user who has been influenced by or accepted new ideas. Users who have not been influenced by the idea yet are inactive nodes. An active node cannot switch to an inactive node, however, the vice-versa is possible. The edges in the network are assigned probabilities that correspond to the possibility of an active node activating its neighboring inactive nodes. 

\subsection{Community Structure}
Community structures often convey prominent features in most complex social networks. They indicate a part of a network with similar characteristics and behavior. They are characterized by the fact that the connections inside a community are noticeably denser than those between different communities \citep{girvan, CBIM}. Communities in social networks can exist due to various reasons, including but not limited to location, common interests, or occupation. 

\begin{figure}
	\centering
		\includegraphics[scale=.5]{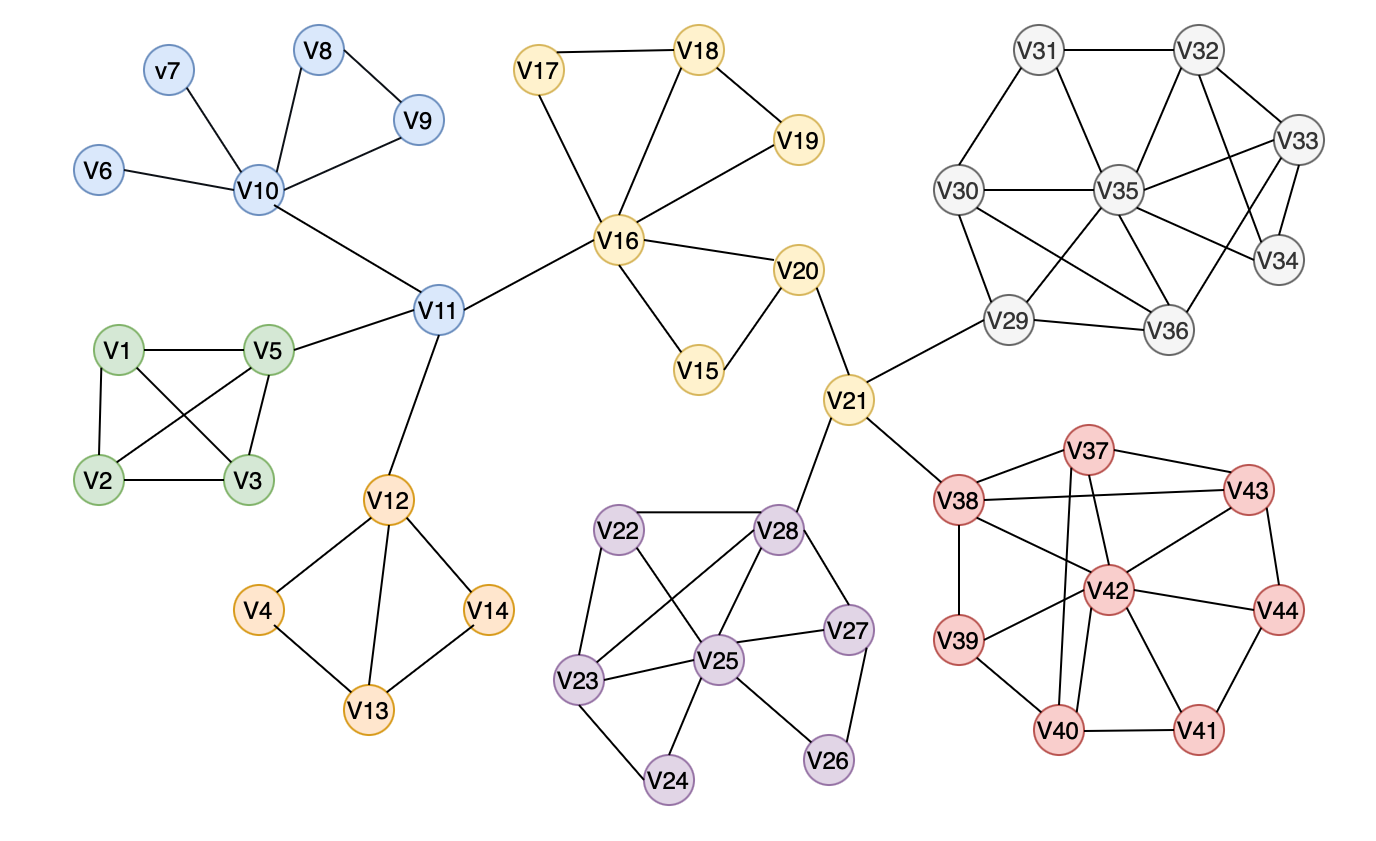}
	\caption{Illustration showing the detection of communities using the Louvain algorithm.}
	\label{communityDetection}
\end{figure}

Detecting communities in complex networks makes it easier to study and analyze the underlying features in the network. It further provides insight into the functioning of systems that are represented by social networks. For example, in metabolic networks, communities represent cycles and pathways. Another instance is the existence of communities in a citation network, representing the group of research papers on the same topic or citing the same article. Community structures are also found in the network of web pages, where each community corresponds to a group of links on a similar topic. Another prominent use of community structures in graphs is that communities often have properties that greatly differ from the standard characteristics of the entire network. In such cases, analyzing only these general characteristics can result in the negligence of important and relevant features. For example, in a social network, both sociable and reserved individuals might exist. Figure \ref{communityDetection} shows communities that are identified in social networks using the Louvain algorithm.

We employ community structures in our approach in the following way:
\begin{itemize}
    \item First, the graph is divided into its constituent communities using the Louvain algorithm \citep{Blondel_2008} for community detection. It is an efficient approach for community detection previously used in a variety of IM works \citep{Huang2019,Shang2018,Gong2016}.
    \item Then, the communities detected by the Louvain algorithm are detached from each other by eliminating the connections between different communities. 
    \item We define a threshold, and select only those communities larger than the defined threshold. These communities are known as significant communities.
    \item We then assign a seed node budget to each significant community, depending on their size. 
    \item The communities are input into the DHHO algorithm to obtain k most influential seed nodes.
\end{itemize}

\subsection{Influence Estimator}
The Local Influence Estimator (LIE) function is used to approximate the spread of influence in a two-hop area around a set of nodes $S$. The set of seed nodes that achieves the maximum value for the LIE function is the optimal seed node set. For the maximization task, the Harris' Hawks optimization algorithm is employed, which has not been used before for the problem of influence maximization. The Three Degree Theory states that a person's influence over his/her's friends gradually dissipates over 3 degrees of separation. Thus, a node has significant influence only over a three-hop area around it. According to \citet{Pei2015}, a node's influence in its three-hop area is a valid indicator of its influence over the entire network. In our study, we use the Local Influence Estimator (LIE) function, which is a two-hop area influence estimator developed by \citep{Gong2016}. It helps quantify the influence spread for a node set $S$ in its two-hop surroundings. It is calculated as shown in Eq. \ref{Eq:LIE}.

\begin{flalign} \label{Eq:LIE}
LIE(s) & = \sigma_0(S) + \sigma_1^*(S) + \tilde{\sigma_2}(S)\\      \nonumber
     & = k + \sigma_1^*(S) + \frac{\sigma_1^*(S)}{|N_S^{(1)}\backslash S|} \sum_{u\in N_S^{(2)}\backslash S} p_u^*d_u^*\\    \nonumber
     & = k + (1 + \frac{1}{|N_S^{(1)}\backslash S|} \sum_{u\in N_S^{(2)}\backslash S} p_u^*d_u^*)\sigma_1^*(S)\\     \nonumber
     & = k + (1 + \frac{1}{|N_S^{(1)}\backslash S|} \sum_{u\in N_S^{(2)}\backslash S} p_u^*d_u^*)  \sum_{i\in N_S^{(2)}\backslash S} (1 - \prod_{(i,j)\in E,j\in S}(1 - p_{ij}))
\end{flalign}
where $\sigma_0(S)$ represents the initial set of seed nodes, $\sigma_1^*(S)$ and $\tilde{\sigma_2(S)}$ represent the expected one-hop and two-hop information propagation of the node set $S$. $p_u^*$ represents the diffusion model's propagation probability. $d_u^*$ represents the number of edges of node $u$ with nodes in $N_S^{(1)}$ and $N_S^{(2)}$.
We achieve the maximization of the above objective function the Harris-Hawks optimization algorithm. Thus, IM is transformed into an objective function optimization problem.

\subsection{Harris' Hawks Optimization}
Harris' Hawks Optimisation (HHO) \citep{HHO} algorithm is a novel gradient-free, population-based meta-heuristic approach applicable to a variety of optimization tasks subjected to a proper formulation of the objective function. The Harris’ Hawk Optimisation (HHO) is inspired by the explorative nature of the prey, as well as different attacking strategies like the surprise pounce of Harris' hawks. This optimization technique can be divided into two phases: the explorative phase and the exploitative phase.

In the exploration mechanism, the hawks use their powerful eyes to detect and track the prey. However, occasionally the prey is not visible to the hawks. Hence, the hawks patiently wait and observe the deserted sites to detect their prey. Once their prey is detected, they switch to their exploitation phase and plan their pounce accordingly. The prey (usually a rabbit) represents the best candidate solution, while the hawks are the other candidate solutions. The Hawks strategize their exploitation technique depending upon the escaping energy and jump strength left with the prey. This energy and jump strength can be modeled as shown in Eq. \ref{escapingEnergy} and Eq. \ref{jumpStrength}, respectively. The value of $E$ decreases upon increasing the number of iterations increases. $E$ indicates the escaping energy of the prey, $J$ indicates the jump strength, $T$ represents the maximum number of iterations, and $E_0$ is the initial energy state. The exploitation phases can be broadly categorized into two varieties of pounces: soft besiege and hard besiege. The former focuses on following the prey until it has run out of its escaping energy and then finally encircling and attacking the prey, while the latter aims at ambushing the prey with a surprise pounce once it is exhausted.

\begin{equation}\label{escapingEnergy}
    E = 2E_0(1-\frac{t}{T})
\end{equation}
\begin{equation}\label{jumpStrength}
    J = 2\times(1-rand())
\end{equation}

\section{Proposed Methodology}
\begin{figure}[t!]
	\centering
		\includegraphics[scale=.75]{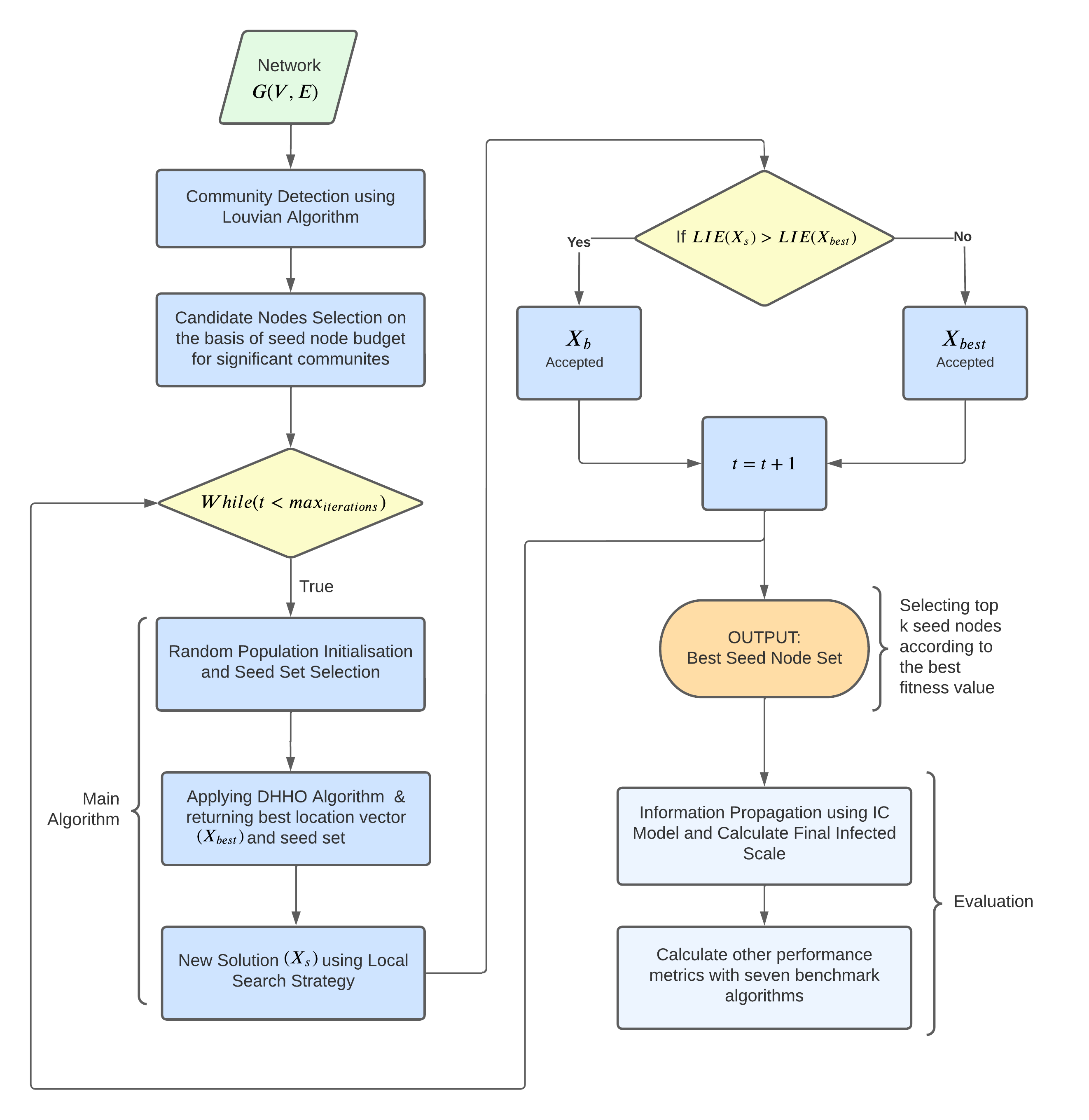}
	\caption{Flowchart representation of the steps involved in our proposed DHHO approach.}
	\label{flowchart}
\end{figure}

We introduce Discrete Harris’ Hawks Optimisation (DHHO), our proposed approach for Influence Maximization. DHHO skillfully combines discretized Harris’ hawks, community structures, and a neighbor scout strategy to extract optimal and most effective seed nodes for maximum information propagation in a given social network. We employ the Louvain algorithm to initialize several communities in social networks, which aids in information propagation and reduces the execution time complexity in our proposed approach. DHHO further utilizes an adapted random probability-based initialization strategy to increase the convergence of the hawk population at the early stages of Harris' hawk manipulation. Figure \ref{flowchart} shows the flow of process in our proposed DHHO framework for IM.

\subsection{Proposed Algorithm}
IM involves finding a group of initial spreader nodes in a network, which can maximize the influence spread across the network. This is an optimization problem, which we aim to solve using the Local Influence Estimator (LIE) function. The LIE function enables us to gauge the expected information diffusion for any set of nodes in a given network. Swarm and population intelligence-based algorithms utilize efficient evolution mechanisms for optimizing the objective function. Here, we propose a discretized form of Harris' hawk optimization (DHHO) algorithm for IM. Since the solution space of the IM task is discrete, the unmodified Harris’ hawk framework cannot be used for IM, as it is suitable only for continuous problems. Thus, we propose the redefinition of the solution vector in a discrete form. The rules for initializing the location vector of Harris' hawks are given below in Eq. 6.

\begin{equation} \label{longEqn}
    X(t+1)=
    \begin{cases}
        X_{rand}(t) - r_1|X_{rand}(t)-2r_2X(t)|,& \text{if } q\geq 0.5\\
        (X_{rabbit}(t)-X_m(t))-r_3(LB+r_4(UB-LB)),              & \text{if } q<0.5
    \end{cases}
\end{equation}
\begin{equation}\label{eq2}
    X_m(t) = \frac{1}{N}\sum_{i=1}^N X_i(t)
\end{equation}

Here, $q$ is the probability of selecting different perching strategies for the hawks, $X(t + 1)$ denotes the position vector of hawks in the upcoming iteration $t$, $X_{rabbit}(t)$ is the rabbit's position (best solution), and $X(t)$ is the current position vector of hawks. The variables $r_1$, $r_2$, $r_3$, $r_4$, and $q$ are randomly chosen and have values between 0 and 1, which are updated in each iteration. $LB$ and $UB$ represent the lower and upper bounds of the variables. $X_{rand}(t)$ is a randomly selected hawk from the current populace, and the $X_m(t)$ is the mean position of all of the hawks from the current populace.

There are four different possible exploitation techniques. Let there be a variable $r$ indicating whether a prey escapes the hawks. ($r<0.5$) depicts the possibility of the prey escaping the hawks, while ($r\geq0.5$) depicts the possibility of the hawks catching the prey.
\begin{itemize}
    \item If the energy level $|E|\geq0.5$ and $r\geq0.5$, then the prey has sufficient energy to escape the hawks by some random, misleading movements. In this strategy, the Harris’ hawks gather around the prey till it gets exhausted and then perform a pounce. This is called soft besiege. 
    \item If the energy level $|E|<0.5$ and $r\geq0.5$, then the prey doesn't have sufficient energy to escape and is exhausted. The Harris' hawks simply perform a pounce to catch the prey. This strategy is called hard besiege. These two approaches are demonstrated mathematically below in Eq. 8 and Eq. 9.
    
\begin{equation}\label{eq3}
    X(t+1) = \Delta X(t) - E|JX_{rabbit}(t)-X(t)|,
    \Delta X(t) = X_{rabbit}(t) - X(t) 
\end{equation}
\begin{equation}\label{eq4}
    X(t+1) = X_{rabbit}(t) - E|\Delta X(t)|
\end{equation}
    
    \item If the energy level $|E|\geq0.5$ and $r<0.5$, then the prey has sufficient energy to successfully escape the hawks by some leapfrog movement, which can be depicted mathematically using the Levy Flight function. In this condition, the Harris' hawks try to aim at the prey perfectly but do not manage to catch it. It makes multiple rapid dives considering the deceptive motions of the prey. This strategy is called soft besiege with rapid progressive dives.
    \item If the energy level $|E|< 0.5$ and $r<0.5$, the prey is exhausted and has no energy left to escape. In this condition, a hard besiege is constructed before the surprise pounce to catch and kill the prey, at the same time decreasing the distance of the average location of the prey. This strategy is called hard besiege with rapid progressive dives. Eq. \ref{eq5} and Eq. \ref{eq6} correspond to both besieges, where $S$ is a random vector of dimension $(1,D)$, while $LF(D)$ represents the Levy flight function, computed using Eq. \ref{LV}. Eq. \ref{eq8} signifies the updated location vector for both besieges with rapid progressive dives using $Y$ obtained from Eq. \ref{eq5} and \ref{eq6} (for soft and hard besieges respectively), and $Z$ from the Eq. \ref{eq7}.

\begin{equation}\label{eq5}
    Y = X_{rabbit}(t)-E|JX_{rabbit}(t)-X(t)|\
\end{equation}
\begin{equation}\label{eq6}
    Y = X_{rabbit}(t)-E|JX_{rabbit}(t)-X_m(t)|
\end{equation}
\begin{equation}\label{eq7}
    Z = Y + S\times LF(D)
\end{equation}
\begin{equation}\label{eq8}
    X(t+1)=
    \begin{cases}
        Y,& \text{if } F(Y)< F(X(t))\\
        Z,& \text{if } F(Z)< F(X(t))
    \end{cases}
\end{equation}
\begin{equation}\label{LV}
    LF(x) = 0.01\times\frac{u\sigma}{|v|^{(1/\beta)}}, \sigma = \left(\frac{\Gamma(1+\beta)\times sin(\frac{\pi \beta}{2})}{\Gamma(\frac{1+\beta}{2})\times\beta\times2^{(\frac{\beta-1}{2}})}\right)
\end{equation}
\end{itemize}

Our proposed method follows a unique discretization procedure for influence maximization, involving the judgment of candidate nodes on the basis of their selection probability. According to \citet{seikhahmadi}, the nodes with degree 1 have minimal possibility of being selected as seed nodes. Thus, to reduce the computational cost of the DHHO algorithm and to make sure it does not get trapped with lesser-degree nodes in a loop, only nodes with degrees greater than one are considered possible seed nodes.

In DHHO, each hawk (solution) has two properties: a location vector and a corresponding seed set. The position of hawk $i$ is represented using a vector with elements. The $j^{th}$ index conveys the possibility of the node being selected as a seed node. The corresponding seed set of hawk $i$ is shown as $S_i$. It contains $k$ nodes with the highest probability of becoming the final seed nodes. The $k$ nodes are then selected according to their decreasing probability. The optimal solution is chosen as the $k$ seed set with the highest probability. This solution is then used to calculate the value of other hawk solutions $X_i$. Once $X_i$ is updated according to the rules of Harris' hawk optimization, then their corresponding seed set values are ranked with respect to their fitness value. We use a neighbor scout strategy to choose an alternate seed set, which is then compared with the main solution. The best seed set is then selected as the global best solution. This process is repeated after each iteration, till the maximum number of iterations is reached.

\begin{algorithm}
\caption{Pseudocode of Discrete Harris' Hawks Optimisation}
\label{GWIM}
\begin{algorithmic}[1]
\State \textbf{Input:} Undirected Graph G = (V,E), seed set size k, population size N, number of maximum iterations T
\State \textbf{Output:} S \Comment{Set of k nodes as initial seedset}
\State Initialize E, J, and $|V'| = \{v_i \in V | d_i > 1$\}
\State Initialize $X_1$ to $X_n$ randomly and determine seed set $S_i$ corresponding to each $X_i$
\State Calculate fitness value for each $S_i(i=1,2,...N)$
\State Assign the best solution as $X_{best}$ (also known as $X_{rabbit}$) according to the fitness value
\State Set t = 0
\While{($t<T$)}
    \For{each hawk i (except $X_{best}$)}
        \State Update E and J using Eq. \ref{escapingEnergy}, \ref{jumpStrength}
    \If{$(|E|\geq1)$}   \Comment{Exploration phase}
       \State Update the location vector using Eq. \ref{longEqn}
    \EndIf
    \If{$|E|<1$}    \Comment{Exploitation phase}
       \If{($r\geq0.5$ and $|E|\geq0.5$)}   \Comment{Soft besiege}
            \State Update the location vector using Eq. \ref{eq3}
        \ElsIf{($r\geq0.5$ and $|E|\leq0.5$)}   \Comment{Hard besiege}
            \State Update the location vector using Eq. \ref{eq4}
        \ElsIf{($r<0.5$ and $|E|\geq0.5$)}      \Comment{Soft besiege}
            \State Update the location vector using Eq. \ref{eq5},\ref{eq7},\ref{eq8}
        \ElsIf{($r<0.5$ and $|E|<0.5$)}     \Comment{Hard besiege}
            \State Update the location vector using Eq. \ref{eq6},\ref{eq7},\ref{eq8}
        \EndIf
    \EndIf
        \State Update $X_i$ and $S_i$ according to the values of E and q
    \EndFor
    \State Calculate the best location vector
    \State Calculate fitness value for each $S_i$ 
    \State Find new solution using neighbor scout strategy as $F_{new}$
    \If{($F_{new}>Fitness_i$)}
        \State $g_{best}$ = new solution
    \EndIf
    \State $t = t+1$
\EndWhile
\State \textbf{Return} $X_{best}$

\end{algorithmic}
\end{algorithm}

The framework for our DHHO algorithm for IM is shown in Algorithm \ref{GWIM}. In line 4, the primary solution vectors are generated according to the initialization strategy. A random position function generates each random solution. In line 5, the LIE function is used to compute the fitness value of each solution. The optimal solution $X_{best}$ is determined using the fitness value of the solutions. In lines 8-34, the operations are iterated for the maximum number of iterations to maximize the fitness value. In each iteration, by using the update strategy and seed set selection function, the positions of the hawks are updated iteratively. Seed set $S_i$ is determined for each hawk in line 25. Line 28 computes the fitness value for each $S_i$, and based on the fitness value, the best hawk particle is chosen. In the end, a neighbor scout strategy is used to find nearby solutions which can perform better than the updated solutions. If the solution generated by the neighbor scout method has a greater fitness value than the original solution, then the former solution is picked accordingly. It aids in preventing being trapped in the local optimum solution.

\subsection{Initialization}
In our proposed DHHO algorithm, we employ an adapted random position initialization function \citep{Zareie2020} to generate the initial solution for the hawk $i$ and its associated seed set. The pseudo-code of this function is shown in Algorithm 2. Each node $V_j$ ($V_j \subset V$) in the set $V$ is accessed, and the position $X_{ij}$ is assigned with the help of a random variable $r$ and the degree of node $V_j$, as shown in lines 3-5. Then, the corresponding seed set is initialized for each node $V_j$, in lines 8-10. The final location vector and seed set values are returned.

\begin{algorithm}
\caption{Random position function}
\label{randomPositionFunction}
\begin{algorithmic}[1]
\State \textbf{Input:} Undirected Graph G = (V,E), V', the seed set size k
\State \textbf{Output:} $X_i$, the position of hawk i, and $S_i$, corresponding seed set
\For{j = 1 to $|V'|$}
    \State $r = rand(1...d_j)$
    \State $X_{ij} = r/max(d)$
\EndFor
\State Set $S_i=\{\}$
    \For{i = 1 to k} 
        \State Find the next maximum $X_{ij}$
        \State Add $v_j$ to $S_j$
    \EndFor
    \State Return $X_i$ and $S_i$
\end{algorithmic}
\end{algorithm}

\subsection{Update Strategy}
Let us take an example to explain the working of the update strategy of the location vector for DHHO clearly. Assuming that all the necessary constants are already defined, let us consider a network in which nodes 1, 5, 6, 9, 13, 15, 20, 26, 29, and 33 have degrees greater than 1. We have set the seed set number $k$ as 3 ($k = 3$) while the nodes of the graphs chosen are $V' = \{1, 5, 6, 9, 13, 15, 20, 26, 29, 33\}$ and $|V'| = 10$. Consider the position vector assigned for each hawk $i$ has been initialized according to the random initialization strategy as:

\begin{table}[h!]
    \centering
    \begin{tabular}{|c|c|c|c|c|c|c|c|c|c|c|}
    \hline
        Node ID & 1 & 5 & 6 & 9 & 13 & 15 & 20 & 26 & 29 & 33\\ \hline
        $X_{ij}$ & 0.436 & 0.213 & 0.121 & 0.456 & 0.746 & 0.589 & 0.322 & 0.234 & 0.886 & 0.055\\ \hline
    \end{tabular}
    \label{tbl2}
\end{table}

For each $X_{ij}$, the $i^{th}$ index represents the population size, and $j^{th}$ node represents the chance of the node in $V$ to be selected as the seed node. For example, the chance of the $1^{st}$ node is 0.436, while it is 0.213 for the $5^{th}$ node and 0.161 for the $6^{th}$ node, and so on. The $k$ nodes with the highest probability of being selected as the corresponding seed set, that is, $S_i = {13,15,29}$.

Let us consider the $max_{iter}$ = 10, and we have to update the location of hawk $i$ for the $6^{th}$ iteration, i.e., $t$ = 6. The value $E$ (refer to Eq. \ref{escapingEnergy}) is calculated as $E = E_0(1- \frac{6}{10})$ and $J = (1 - rand())$, where $E_0 = (2\times rand() - 1)$. A random value is drawn from the domain $[0, 1]$ as $r$.


Let us suppose $r = 0.75$, $E = 0.2$ and $J = 0.25$. For this condition, $|E|< 1$, $|E|<0.5$ and $rand() > 0.5$, hard besiege is activated. Calculating the value of X, we get $X(t) = X_{best} - E*(J*X_{best} - X(t))$. 

$X(t)$ turns out to be 0.345. Similarly, this process is carried out till the $max_{iter}$ is reached for all the hawks. Then, the seed set and fitness value are calculated for each hawk in the population accordingly. If the fitness value of the current solution exceeds the ${X_{rabbit}}$ solution, they get replaced with the best global solution.

\subsection{Neighbor Scout Strategy}
We present a novel Neighbor Scout Strategy (NSS) that searches for a better solution vector by replacing a few nodes with their neighboring nodes. This strategy aims to find a new solution vector near the vicinity of the current location vector in the search space. It ensures that the solution does not get trapped in the algorithm's local optimum solution of the search space. The nodes are replaced with their neighbor nodes only if the size of their one-hop neighbor is large enough. The original Harris' hawks algorithm tends to have a slower convergence speed for computationally demanding optimization tasks. Thus, this strategy further assists in accelerating the convergence speed of the proposed algorithm. The neighbor scout strategy (NSS) used in our proposed DHHO is illustrated in Algorithm \ref{localSearch}.

\begin{algorithm}
\caption{Neighbor Scout Strategy ($x_i$,G, L)}
\label{localSearch}
\begin{algorithmic}[1]
\State \textbf{Input:} Graph G = (V,E), position $x_i$.
\State $d_{x_i} \gets Degree (x_i)$
\State $X_i' \gets Order(x_i,d_{x_i})$
\State $N \gets Neighbours(X_i)$
\For{each element $X_{ij}' \in X_j'$}
    \For{each node $N_i$ in $N^{(1)}_{X'_{ij}}$} 
    \If{{$r>0.5$ and $len(N^{(1)}_{X'_{ij}})> L$}}
        \State $X'_i \gets Replace(X'_{ij},N_i)$
        \If{$LIE(X'_i)>LIE(x_i)$}
            \State $x_i \gets X'_i$
        \EndIf
    \EndIf
    \EndFor
\EndFor
\State \textbf{Output:} The new position $x_i$
\end{algorithmic}
\end{algorithm}

First, the current solution of nodes is sorted in ascending order based on their degree of centrality. This order ensures that the nodes of lower influence get replaced by more influential nodes in their neighborhood (as shown in lines 1-2). Each node is accessed iteratively and replaced by a node in its one-hop neighbor depending upon a probability $r$. The nodes are only replaced if the size of their neighborhood is bigger than a given threshold $L$, then the LIE fitness value for the whole solution vector is calculated. If the $LIE$ value of the replaced solution vector is greater than the $LIE$ value of the original vector, then the original node gets replaced (demonstrated in lines 5-8).

\section{Datasets, Baseline Models and Performance Metrics}

\subsection{Datasets}

We evaluate our DHHO approach on six social network datasets of varying domains and sizes. Table \ref{Tab:Datasets} provides various details about the datasets like the number of nodes, edges and communities detected by Louvain's algorithm. They are described below:
\begin{itemize}
    \item Wiki-Vote \citep{wikivote}: This dataset contains voting information from Wikipedia till January 2008. Each node signifies a user. Directed edges between node $u$ and node $v$ indicate that person $u$ voted for person $v$.
    \item Power-law Cluster Graph (PCG) \citep{PCG}: PCG is a technique used for the generation of random graphs using approximate average clustering and power-law degree distribution. Parameters $n$, $m$, and $p$ are required to generate a graph, where $n$ represents the number of nodes in the graph, $m$ denotes the number of edges per node in the graph, and $p$ represents the probability of a triangle is created upon adding a new edge to the graph.
    \item Email-univ \citep{email-univ}: This dataset contains a network obtained from a research institution in Europe, using email data from October 2003 to May 2005. The nodes signify the users, while the directed edges in the graph denote the flow of emails.
    \item Jazz \citep{jazz}: The Jazz dataset contains a network denoting collaborations in the jazz music community collected in 2003. The nodes signify Jazz musicians, while edges represent collaborations between two Jazz musicians. 
    \item Hamsterster \citep{hamsterster}: The hamsterster dataset represents a network obtained from hamsterster.com. The nodes represent users on the website, while edges signify a friendship between the nodes.
    \item P2P-Gnutella \citep{gnutella}: This dataset contains snapshots of the Gnutella file-sharing network. The nodes indicate a host in the network, while edges signify connections between the hosts.
\end{itemize}

\begin{table}[t!]
    \centering
    \caption{The different datasets used for experimentation along with the various characteristics such as the Number of Nodes, Edges, and Communities detected by the Louvain Algorithm}
    \begin{tabular}{|c|c|c|c|}
    \hline
        Dataset & Nodes (n) & Edges (m) & Communities \\ \hline
        \makecell{Wiki-Vote \citep{wikivote} \\ \citep{wikivote2}} & 889 & 2914 & 40\\ \hline
        PCG \citep{PCG} & 2000 & 9963 & 21 \\ \hline
        Jazz \citep{jazz} & 198 & 2,742 & 40 \\ \hline
        Email-univ \citep{email-univ} & 1100 & 5500 & 11 \\ \hline
        Hamsterster \citep{hamsterster} & 2,400 & 16,600 & 169 \\ \hline
        p2p-Gnutella \citep{gnutella} & 8,846 & 31,839 & 23 \\ \hline
    \end{tabular}
\label{Tab:Datasets}
\end{table}

\subsection{Baseline Approaches}
We compare DHHO with seven existing baseline approaches for IM over five evaluation metrics for a comprehensive comparative analysis. A brief overview of these approaches is shown below.

\begin{itemize}
    \item Extended Neighborhood Coreness (ENC) \citep{ENC}: ENC is a centrality measure used for IM. It is based on the K-shell decomposition \citep{kshell}. It extends over K-shell decomposition to a two-hop measure by taking the sum of both one and two-hop measures.
    \item Gateway Local Rank (GLR) \citep{GLR}: GLR is an extension of the closeness centrality measure and simplifies it by reducing the search set to the local and gateway nodes.
    \item PageRank (PR) \citep{Brin1998}: PageRank measures the significance of a given node in the network based on its outgoing degree and damping factor.
    \item H-index (HI) \citep{HI}: H-index is a node centrality measure that considers the neighborhood measures, the quality, as well as the quantity of the neighboring nodes simultaneously.
    \item Discrete Particle Swarm Optimisation (DPSO) \citep{Gong2016}: DPSO involves the use of the discretized Particle Swarm Optimization (PSO) \citep{PSO} for IM with a local search strategy and degree-based heuristic initialization.
    \item Discrete Bat Algorithm (DBA) \citep{batAlgo}: DBA meta-heuristic approach is used for Influence Maximization using a probabilistic local search strategy and random walk strategy.
    \item Efficient Discrete Differential Evolution (E-DDE) \citep{EDDE}: E-DDE was used for Influence Maximization with community structures and population-based initialization.
\end{itemize}

\subsection{Performance Metrics}
\begin{itemize}
    \item Final Infected Scale (FIS): FIS is the ratio of the infected nodes after the simulation ends and the total nodes in the graph. The activation of nodes is caused due to the activation probability working in the favor of the nodes. The initial spreader fraction plays a critical role in the final number of infected nodes in the graph. Generally, a higher spreader fraction leads to a higher FIS.

    \item LIE and log(LIE) function vs Spreader Fractions: The LIE function has a prominent contribution in evaluating the FIS in a social network. The LIE function value represents the fitness value of the objective function for IM. Usually, the LIE fitness value has a large magnitude hence, we use a natural logarithm (log)function for convenience. The greater the fitness value of the objective function, the higher is the influence of the seed node. LIE and log(LIE) function values are plotted against each spreader function, logically a higher spreader fraction leads to a higher LIE fitness value. The LIE and log(LIE) values indicate the accuracy of the approach.	
    
    \item Activation Probability (P) vs Final Infected Scale (FIS): Activation Probability (P) plays a crucial role in the total final spread. Typically, as P increases, the FIS value also increases. FIS indicates the number of infected nodes at the end of the spread time. Increasing the activation probability means increasing the chances of activating more nodes for information propagation, hence increasing the final influence diffusion.

    \item Running Time: In order to compare our proposed approach with other baseline approaches, we also take into account the amount of time it takes to run the whole algorithm and rank nodes based on their ability to influence the network. The time consumption was evaluated while running all approaches in a Jupyter Environment, on a system with a Tesla K80 GPU and 12GB RAM available to the environment.

\end{itemize}

\section{Experimental Results and Analysis}

\begin{figure}[t]
	\centering
		\includegraphics[scale=.2]{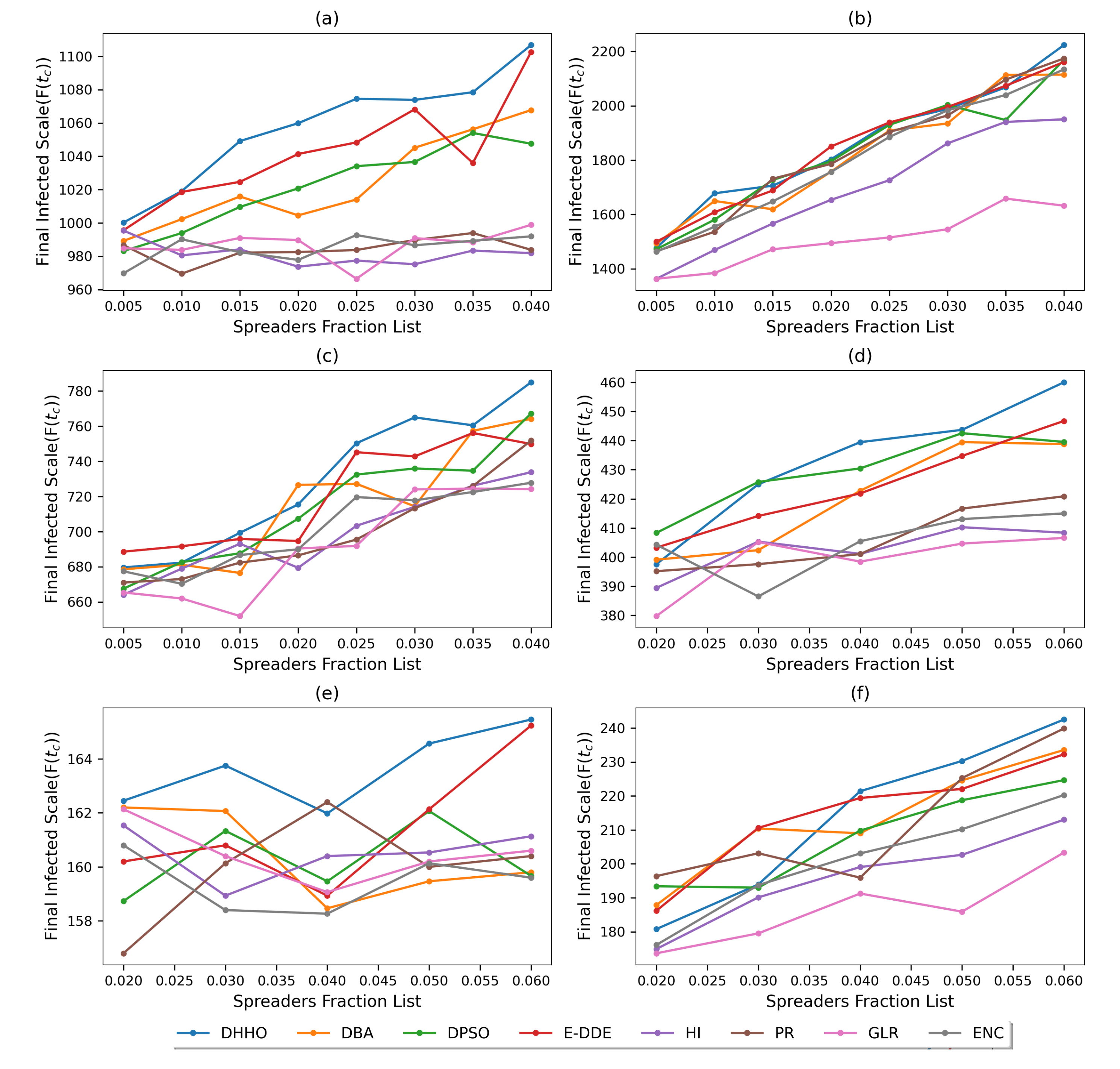}
		
	\caption{Final infected scale plots for various values for initial spreader fraction on the (a) hamsterster dataset (b) P2P-gnutella dataset (c) PCG dataset (d) Email-univ dataset (e) Jazz dataset (f) Wiki-vote dataset.}
	\label{finalInfectedScale}
\end{figure}

\subsection{Final Infected Scale (FIS)}

The FIS plots the final infection at the end of the simulation corresponding to the various values of the spreader fraction. We ran the Independent Cascade (IC) simulation 50 times for our proposed DHHO approach as well as for other competing approaches on six publicly available datasets and plotted the average values for various values of the spreader fraction. The spreader fraction value lies in the set {0.005, 0.01, 0.015, 0.02, 0.025, 0.03, 0.035, 0.04} for larger datasets (nodes > 2000), and in the set {0.02, 0.03, 0.04, 0.05, 0.06} for smaller datasets. For smaller datasets, we have taken relatively larger spreader fractions because smaller values lead to a very small absolute number of initial spreaders in smaller datasets. The activation probability was taken to be 0.1 for all methods and datasets. Figure \ref{finalInfectedScale} shows the plots for the FIS vs Spreader fraction for DHHO and the baseline approaches. The initial spreader fractions are plotted along the x-axis, while the corresponding FIS values are plotted along the y-axis.

\begin{figure}[t]
	\centering
		\includegraphics[scale=.2]{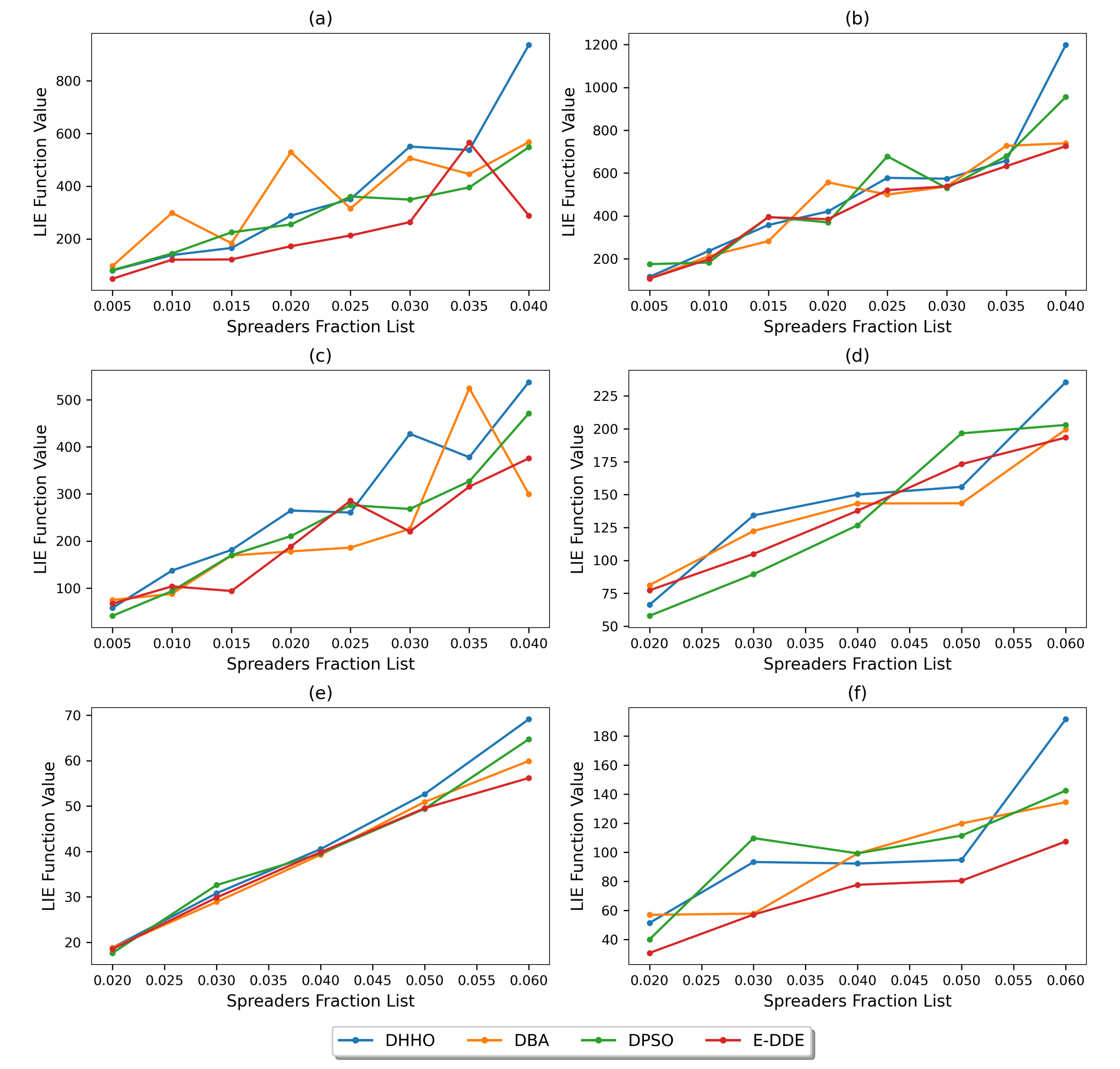}
	\caption{LIE Function Value plots for various values of initial spreader fraction for the (a) hamsterster dataset (b) P2P-gnutella dataset (c) PCG dataset (d) Email-univ dataset (e) Jazz dataset (f) Wiki-vote dataset.}
	\label{LIEfunctionValue}
\end{figure}    

We observed that for most datasets, as the spreader fraction increases, the final infected scale value also increases correspondingly. However, for a few approaches like  PR, GLR, and HI, the final infected scale did not increase continuously upon increasing the spreader fraction for some datasets. Our proposed approach, DHHO, appeared to perform significantly better than the other baseline approaches. Other meta-heuristic approaches like DBA and E-DDE performed well on all datasets too and sometimes outperformed DHHO  for a given value of the spreader fraction. However, in general, DHHO performs noticeably better than other meta-heuristic and heuristic algorithms across all datasets and all values of spreader fraction.
    
\subsection{LIE vs Spreader Fractions and log(LIE) vs Spreader Fraction}

\begin{figure}[t]
	\centering
		\includegraphics[scale=.2]{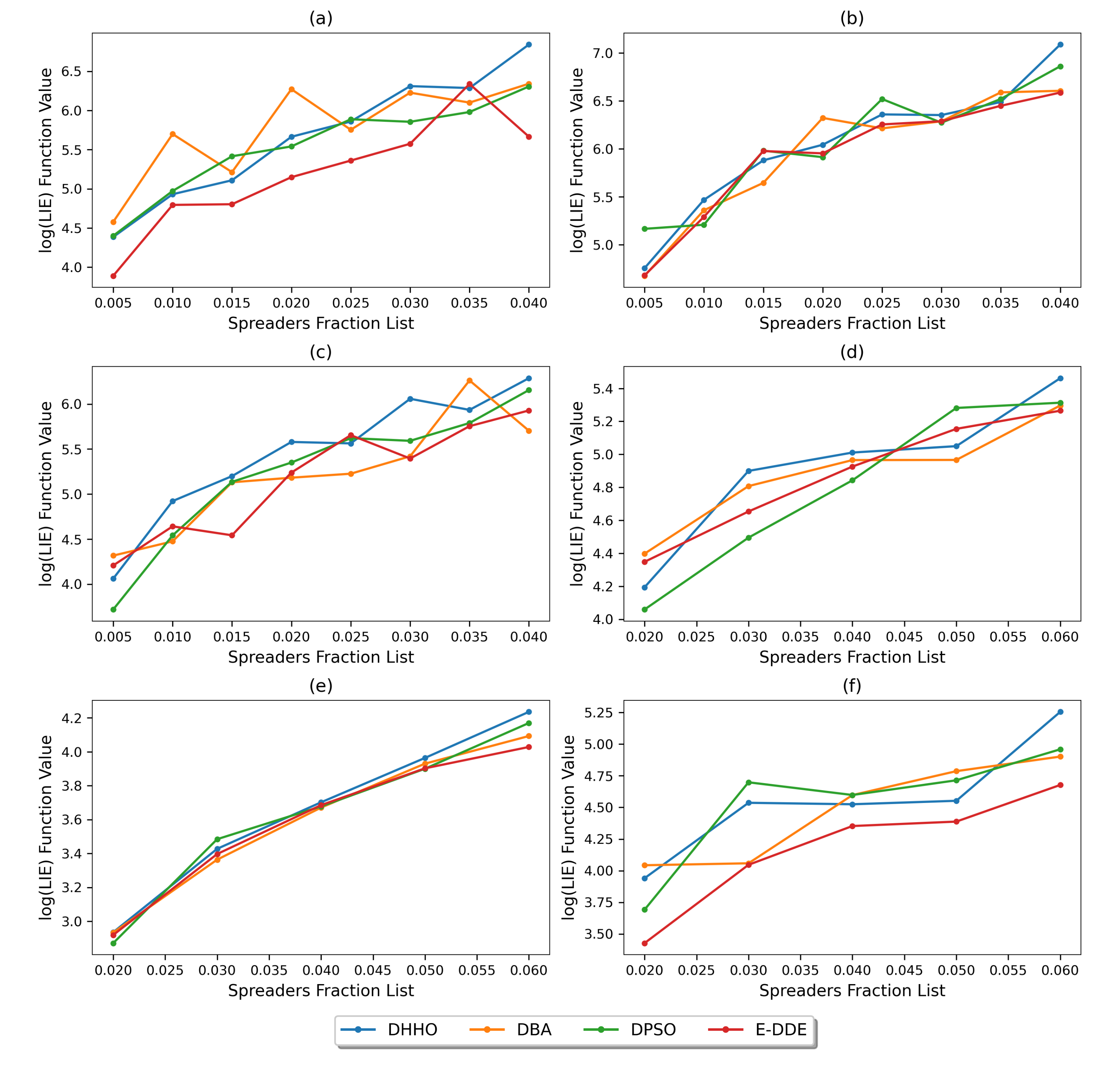}
	\caption{log(LIE) Function Value plots for various values of initial spreader fraction on the (a) hamsterster dataset (b) P2P-gnutella dataset (c) PCG dataset (d) Email-univ dataset (e) Jazz dataset (f) Wiki-vote dataset.}
	\label{log(LIE)}
\end{figure}

We evaluated the performance of the fitness value of the  Local Influence Estimator (LIE) objective function against the different values of the spreader fractions for DHHO and other baseline meta-heuristic approaches. We plotted LIE and log(LIE) Function Values for different values of spreader fraction as shown in Figures \ref{LIEfunctionValue} and \ref{log(LIE)}. The spreader fraction values are plotted along the x-axis and the corresponding LIE function values are plotted along the y-axis. The initial spreader fraction values for datasets having greater than 2000 nodes were chosen from the set {0.005, 0.01, 0.015, 0.02, 0.025, 0.03, 0.035, 0.04}, whereas for smaller datasets with less than 2000 nodes, the initial spreaders fraction were taken from the set {0.02, 0.03, 0.04, 0.05, 0.06}. Once again, for smaller datasets, we have taken relatively larger spreader fractions because small spreader fractions will lead to an insignificant number of initial spreaders in small datasets.

The log(LIE) metric is similar to the LIE function metric but the slope or the curvature of the graph changes significantly. As the spreader fraction increases, the fitness value increases as well, and so does the final influence diffusion. Our proposed approach DHHO performs as well or better than the competing approaches on all datasets. While DBA and DPSO sometimes performed slightly better than DHHO at some value of the spreader fraction for some datasets, DHHO performed more consistently and was noticeably better for higher values of spreader fraction. Thus, we can conclude that DHHO performs as well, if not better than competing novel meta-heuristic approaches for IM in terms of LIE-vs-Spreader-Fraction and log(LIE)-vs-Spreader-Fraction.
    
\begin{figure}[t]
	\centering
		\includegraphics[scale=.2]{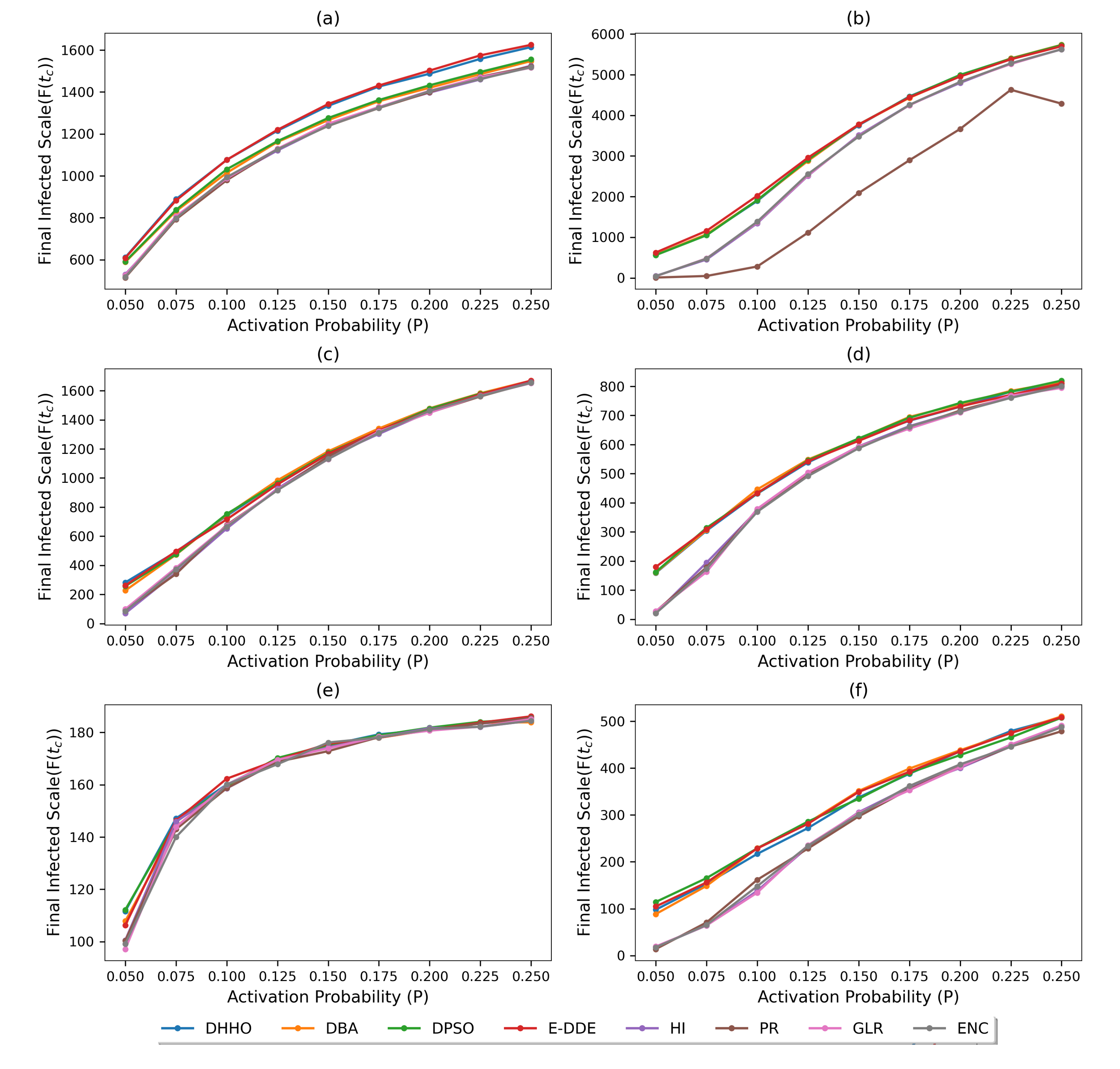}
	\caption{Final Infected Scale plots for different activation probability $(p)$ values on the (a) hamsterster dataset (b) P2P-gnutella dataset (c) PCG dataset (d) Email-univ dataset (e) Jazz dataset (f) Wiki-vote dataset. The initial spreader fraction was set as 0.10.}
	\label{probability}
\end{figure}    
    
\subsection{Activation Probability (P) vs Final Infected Scale (FIS)}

We compared the performance of DHHO with other baseline approaches on the basis of FIS for different values of activation probability (P). We took the different activation probability values in the range {0.05, 0.075, 0.1, 0.125, 0.15, 0.175, 0.2, 0.225, 0.25}. The initial spreader fraction for each case was taken to be 0.10, and for each dataset, the simulation was performed 50 times. Figure \ref{probability} shows the FIS vs P plots for the different methods on all six datasets. The values for the activation probabilities are taken along the x-axis, whereas the Final Infected State is taken along the y-axis. From the given plots, we can observe that as per our expectation, the FIS increased upon increasing the value of P. Our proposed method DHHO and E-DDE perform significantly better than the competing approaches on all datasets. Other methods like DBA and DPSO performed sometimes perform comparably to DHHO, but do not perform better than DHHO at any spreader fraction. HI, PR, GLR, and ENC are noticeably behind the meta-heuristic approaches on all six datasets.
    
\begin{figure}[t]
	\centering
		\includegraphics[scale=.2]{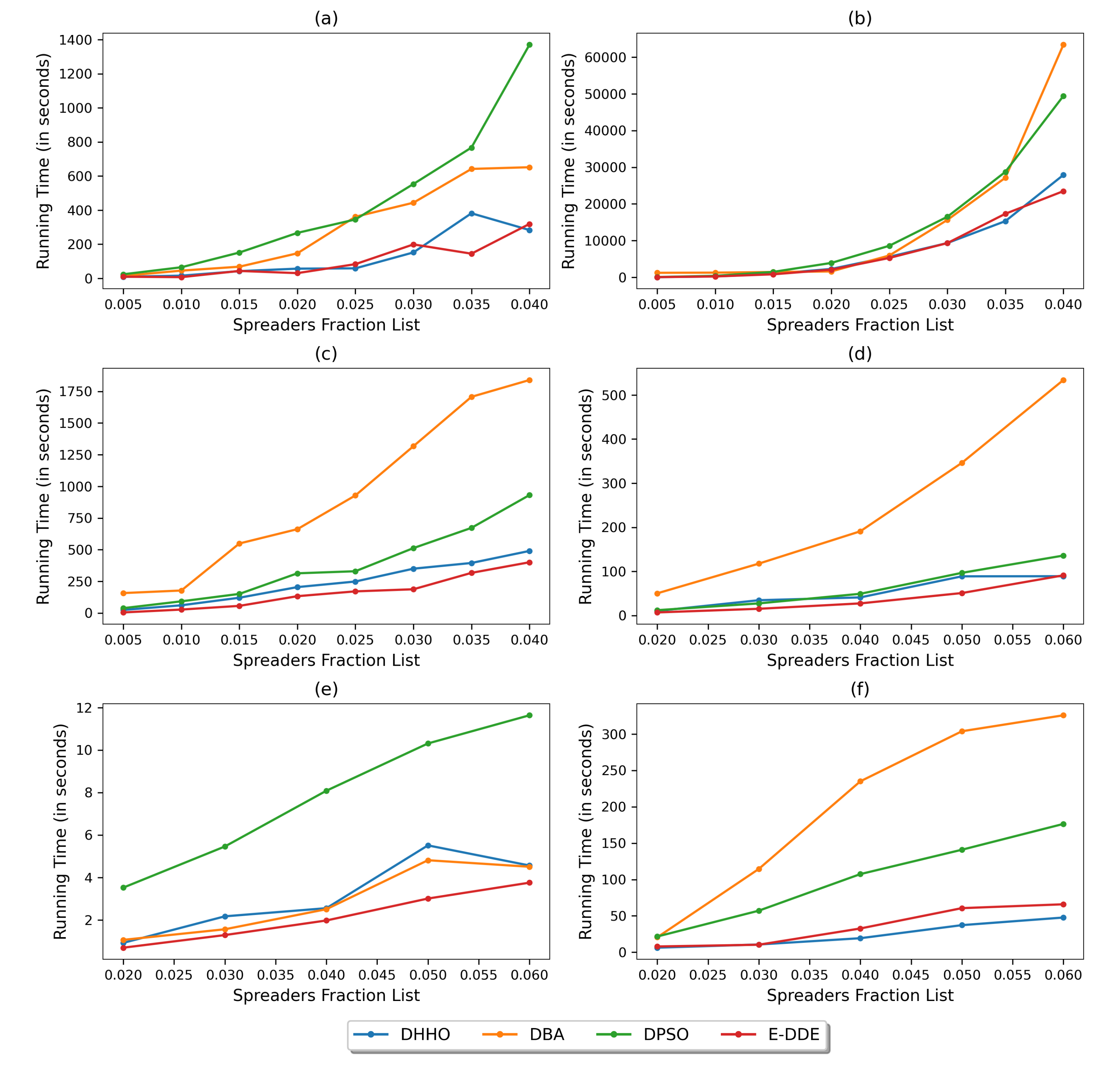}
	\caption{The execution time plots for different methods for ranking the spreaders for the (a) hamsterster dataset (b) P2P-gnutella dataset (c) PCG dataset (d) Email-univ dataset (e) Jazz dataset (f) Wiki-vote dataset.}
	\label{runningTime}
\end{figure}    
    
\subsection{Running Time}

We also evaluated our proposed approach DHHO against the other competing meta-heuristic approaches on the basis of time execution for compiling the final influence diffusion. Figure \ref{runningTime} shows our plots for execution time vs spreader fraction. The various values of the spreader fraction are plotted along the x-axis, while the execution time (in seconds) is plotted along the y-axis. The initial spreader fraction values again for datasets having greater than 2000 nodes were chosen from the set {0.005, 0.01, 0.015, 0.02, 0.025, 0.03, 0.035, 0.04}, whereas for smaller datasets with less than 2000 nodes, the initial spreaders fraction were taken from the set {0.02, 0.03, 0.04, 0.05, 0.06}.

We observed that DPSO and DBA usually have a longer execution time than DHHO and E-DDE. Since DHHO and E-DDE employ community structures and do not employ lengthy functions such as random walk strategy, the execution time for DHHO and E-DDE is significantly lesser than the other meta-heuristic approaches. DHHO and E-DDE are comparable across three datasets, DHHO performs better than E-DDE on one dataset, while E-DDE outperforms all approaches on two datasets in terms of execution time. However, E-DDE does not have the best final infected scale value. Thus, DHHO, with a similar execution time and a higher final influence spread across almost all datasets and spreader fractions, is a better approach when taking into account the holistic performance across different metrics.

\section{Statistical Testing}
To further evaluate the performance of DHHO using statistical tests, we have conducted the Friedman test \citep{Friedman}, which is a non-parametric statistical test used for multiple comparisons (more than two given methods). We have performed this test to determine whether the performance of our proposed method is significantly different than the other approaches for two or more data sets. The Friedman test attempts to detect significant differences based on the ranking of the methods, rather than their errors. It consists of two hypotheses: the null ($H_0$) and the alternate hypothesis ($H_1$). The former states that there are no prominent differences between these algorithms (equality of medians) and the latter insists that the algorithms have significant differences in their median of the population, thus negating the null hypothesis. We have performed the following test on the Final Infected Scale (FIS) vs Spreader Fractions metric. In order to perform the Friedman Statistical Test, we execute the following procedure:

\begin{itemize}
    \item Gather the generated results for each problem pair.
    \item Rank the values in ascending order from 1 (best value) to $n$ (worst value) for each problem $i$ for a particular algorithm $j$.
    \item For algorithm $j$, the average rank for each problem $i$ is calculated using the Eq. \ref{RJ}:
    \begin{equation}\label{RJ}
        R_j = \frac{1}{n} \sum_{i=1}^n(r_i^j)
    \end{equation}
    where $r^j$ is the rank ($1<j<k$) and $R_j$ is the average rank.
    \item Now that all algorithms are ranked according to their priority, compute the Friedman Statistic $F_f$ using the following equation. $F_f$ is based on a Chi-square distribution, with $k-1$ degree of freedom. $F_f$ is computed as shown in Eq. \ref{FF}. In the following equation, $n$ represents number of rows and $k$ represents number of columns ($n = 8$ and $k=6$). 
    \begin{equation}\label{FF}
        F_f = \frac{12n}{k(k+1)} \left[\sum_{j=1}^k R_j^2 - \frac{k(k+1)^2}{4}\right]
    \end{equation}
    \item The previous statistic produced relatively conservative results, which were not desired. Thus, \citet{Iman-Davenport} introduced a new statistic $F_{id}$, as shown below, which follows the $F-distribution$ with the degree of freedom as $k-1$ and $(n-1)(k-1)$, given in Eq. \ref{Eq:FID}:
    \begin{equation}\label{Eq:FID}
        F_{id} = \frac{(n-1)\chi_F^2}{n(k-1)-\chi_F^2}
    \end{equation}
\end{itemize}

Table \ref{Stat} shows the average ranking calculated from the Friedman test for all the approaches used in this paper. DHHO clearly has a better average ranking than the other competing approaches.

\begin{table}[h!]
    \centering
    \caption{Average ranking of algorithm calculated using the Friedman test}
    \begin{tabular}{|c|c|c|}
    \hline
        Sr No. & Algorithm & Average Ranking \\ \hline
        1 & DHHO & \textbf{1.867} \\ \hline
        2 & DBA & 3.375 \\ \hline
        3 & DPSO & 3.250 \\ \hline
        4 & E-DDE & 3.083  \\ \hline
        5 & HI & 6.416\\ \hline
        6 & PR & 5.208 \\ \hline
        7 & GLR & 6.375 \\ \hline
        8 & ENC & 5.667 \\ \hline
    \end{tabular}
\label{Stat}
\end{table}

The unadjusted P-value or Holm P-value, obtained from the Iman-Davenport statistic, corresponding to the performance of DHHO, advocates the rejection of the null hypothesis $H_0$. All of the computed P-values are less than the standard significance level $\alpha = 0.05$. This indicates that there is a significant difference between the performance of the baseline approaches and our proposed approach. Hence, the obtained P-values conclude the negation of null hypothesis $H_0$ but are not appropriate for comparison with different methods. In order to compare these approaches with each other, we must calculate their respective adjusted P-values keeping DHHO as the control algorithm.

Hence, we use adjusted P-values (APVs) for evaluation in order to perform a comparative study between the control method and benchmark methods on a statistical ground. Adjusted P-value provides the correct correlation between these algorithms, taking into account the accumulated family error with respect to the DHHO control algorithm. Moreover, the adjusted P-values can be directed compared with the significance level $\alpha$, which is equivalent to 0.05. In order to calculate the adjusted P-values, a few post-hoc procedures need to be defined. Various post-hoc procedures such as \citet{bonn} and \citet{holland} differ as they adjust the value of $\alpha$ to compensate for multiple comparisons for multiple methods. In this paper, we have used the common Holm's procedure \citep{Holm} to evaluate the respective APVs. The values of Holm P-values and APVs are always sorted in ascending order. The equation for the same is given below, where indices $i$ and $j$ refers to the main hypothesis whose APVs are being computed and different hypothesis in the set respectively. $P_j$ is the P-value for the $j^{th}$ hypothesis. Holm APV is computed as shown in Eq. \ref{Eq:Holm}.

\begin{equation}\label{Eq:Holm}
    Holm APV_i = \min\{v,1\}, \text{where} \,v = \max\{(k-j)p_j:1\leq j\leq i\}
\end{equation}

\begin{table}[h!]
    \centering
    \caption{Adjusted P-values (APVs) using Holm procedure}
    \begin{tabular}{|c|c|c|c|c|}
    \hline
        Sr No. & Algorithm (DHHO is control algorithm) & Z score & Holm p-value & Adjusted p-value \\ \hline
        \makecell{1} & HI & -7.03 & 9.89e-13 & 6.92e-12 \\ \hline
        2 & GLR & -6.97 & 1.56e-12 &  9.41e-12\\ \hline
        3 & ENC & -5.87 & 2.12e-09 & 1.06e-08\\ \hline
        4 & PR & -5.16 & 1.20e-07  & 4.71e-07\\ \hline
        5 & DBA & -2.32 & 1.00e-02 & 2.79e-02 \\ \hline
        6 & DPSO & -2.13 & 1.65e-02 & 3.31e-02\\ \hline
        7 & E-DDE & -1.87 & 3.06e-02 & 4.99e-02\\ \hline
    \end{tabular}
\label{Holm}
\end{table}
The results of the APVs using Holm's procedure are shown in Table \ref{Holm}. The intended APVs are less than the significance level, thus rejecting the null hypothesis. It is clearly observed from the results obtained that DHHO is significantly better than the competing approaches on these statistical tests as well.

\section{Conclusion}

The tremendous popularity of social networks in today's world has made  influence maximization a crucial research problem in today's world. The selection of optimal seed nodes for maximum influence spread is critical for viral marketing. In this paper, we presented a Discretized Harris' Hawks Optimization approach for IM using community structures for optimal selection of seed nodes for influence spread. We employed a neighbor scout strategy algorithm to avoid blindness and enhance the searching ability of the hawks in the DHHO algorithm. We used a random population initialization strategy based on candidate nodes to accelerate the process of convergence of the population. We verified the performance of our approach by evaluating it on six social network datasets across five metrics. Our proposed DHHO approach outperformed the three competing meta-heuristic approaches and the four competing heuristic approaches. Community structures helped enable our approach to perform extremely efficiently and converge in minimal running time compared to other approaches. Statistical tests done on our obtained results further verified the superiority of our proposed DHHO approach compared to other approaches, indicating that the seed nodes selected by DHHO enable a wider propagation of information in social networks as compared to competing approaches.

\bibliographystyle{cas-model2-names}

\bibliography{main}

\end{document}